\def\be{\begin{equation}}
\def\ee{\end{equation}}
\def\bea{\begin{eqnarray}}
\def\eea{\end{eqnarray}}

\def\apr{\approx}
\def\th{\theta}
\def\d#1d#2{\frac{d #1}{d #2}}
\def\dd#1dd#2{\frac{d^2 #1}{d #2^2}}
\def\del#1#2{\frac{\partial #1}{\partial #2}}
\def\ddel#1#2{\frac{\partial^2 #1}{\partial #2^2}}

\def\gev{\,{\rm GeV}}

\def\cmm2{{\,\rm cm^{-2}}}
\def\cm2{{\,{\rm cm}^2}}
\def\cmm3{{\,{\rm cm}^{-3}}}
\def\gcmm3{{\,{\rm g\,cm^{-3}}}}

\def\mpl{{m_{\rm Pl}}}

\def\vev#1{\left\langle #1\right\rangle}

\def\fun#1#2{\lower3.6pt\vbox{\baselineskip0pt\lineskip.9pt
  \ialign{$\mathsurround=0pt#1\hfil##\hfil$\crcr#2\crcr\sim\crcr}}}
\documentstyle[12pt,fleqn]{article}
\begin{document}
\begin{titlepage}
\vspace*{-62pt}
\begin{flushright}
FERMILAB--Pub--96/159--A\\
June 1996
\end{flushright}
\vspace{1.5cm}
\begin{center}
{\Large \bf Fluctuations and Bubble Dynamics in \\First-Order
Phase Transitions}\\ 
\vspace{.6cm}
\normalsize
{Mark Abney$^*$
\vspace{12pt}

{\it 

Department of Physics, The Enrico Fermi  
Institute,\\
The University of Chicago, Chicago, Illinois~~~60637\\ and\\                     

NASA/Fermilab Astrophysics Center,\\
Fermi National Accelerator Laboratory, Batavia, Illinois~~~60510}
\vspace{18pt}
}

\end{center}

\vspace*{12pt}

\begin{quote}
\hspace*{2em}

We numerically examine the effect of thermal fluctuations on a
first-order phase transition in 2+1 dimensions. By focusing on
the expansion of a single bubble we are able to calculate
changes in the bubble wall's velocity as well as
changes in its structure relative to the standard
case where the bubble expands into a homogeneous background. Not only
does the wall move faster, but the transition from the 
symmetric to the asymmetric phase is no longer smooth, even for
a fairly strong transition. We discuss  how these results affect
the standard picture of electroweak baryogenesis.

\vspace*{12pt}


\vspace*{12pt}

\noindent
\small $^*$email: abney@oddjob.uchicago.edu 

\end{quote}

%
\end{titlepage}
%

\baselineskip=24pt

\section{Introduction}
\label{introduction}

Recent numerical and analytical work on weak first-order phase
transitions has shown that there may be significant changes to
the standard theory of phase transitions as a result
of thermal fluctuations \cite{1}--\cite{6}. Instead of a
smooth homogeneous 
background there may be a significant amount of phase mixing due
to the existence of subcritical bubbles. Though not without
controversy \cite{7},\cite{8}, these 
findings suggest we may need to reevaluate the standard theory
of nucleation of critical bubbles because in very weak
first-order phase transitions the standard assumption of only
small amplitude fluctuations breaks down. As the phase transition
increases in strength we expect the role of fluctuations to
diminish until the approximations made for homogeneous
nucleation become applicable. The effects of thermal
fluctuations may extend beyond the regime of nucleation however,
and alter the dynamics of bubbles as they expand. Due to the
complex nature of the system, analytic investigations of this
question would be difficult. On the other hand, numerical
simulations in 3+1 dimensions would be very computationally
extensive. We therefore address the problem of dynamics by undertaking
numerical simulations in 2+1 dimensions. Our primary motivation
is to gain at least a qualitative understanding of how thermal
fluctuations may affect the electroweak phase transition in the
early universe and the consequent ramifications on baryogenesis
at this epoch. 

Work on the topic of electroweak baryogenesis has, in general,
been concentrated in three areas: the form of the effective
potential \cite{9}, the dynamics of the transition
\cite{8},\cite{10}--\cite{12}, and how to calculate
the baryon asymmetry \cite{13}--\cite{18}. These three areas
effectively form a 
hierarchical structure where the means by which one computes the
baryon asymmetry depends on the dynamics of the transition,
which in turn depends on the form of the effective
potential. Though we do not yet know what the true effective
potential looks like, it nevertheless makes sense to investigate
the other two areas by  assuming certain generalities. That is,
we may choose a theory with a scalar order parameter $\phi$
responsible for spontaneous symmetry breaking.
If we assume a potential which may be approximated by 
\be
  V=a\phi^2-b\phi^3+c\phi^4, 
\label{simple.pot}
\ee
where the coefficients are temperature dependent,
we can compute quantities such as the temperature needed to
nucleate bubbles of the 
new phase, the thickness of the bubble walls, and the rate at
which the old phase is converted to new phase as a function of the
coefficients.\footnote{In 
fact, the rate of phase conversion depends on the velocity of
expansion of the bubble walls, which depends not only on the
strength of the transition but also on the interaction of the
wall with the cosmological plasma.} We
require a first-order transition because of the third of
Sakharov's conditions for baryogenesis, the lack of thermal
equilibrium. Because the cooling rate of the universe is
extremely slow at the electroweak transition, depending only on the
rate of expansion, the 
cosmological fluid maintains thermal
equilibrium. However,  by allowing a
bubble of stable vacuum to appear within the metastable vacuum,
a first-order transition gives rise to out of equilibrium
processes in the neighborhood of the bubble wall. It is only within
this confined region where baryogenesis can take place.

The order and strength of the electroweak phase transition is the
subject of 
ongoing research which will only be solved by a calculation of
the effective potential which is valid to all orders of
perturbation theory. In any case we will
assume a potential of the general form of
eq. (\ref{simple.pot}). 
The strength of the transition is
determined by a ratio of the coefficients in
eq. (\ref{simple.pot}), the value of which is temperature and
model dependent. In the minimal standard model for the electroweak
theory the quark, gauge boson and Higgs masses determine the height of
the energy barrier separating the two minima of the effective
potential. If the Higgs mass is
small the transition is strongly 
first-order, whereas  a large Higgs mass results in a transition 
which is at most weakly first-order and possibly second-order. While
most work on phase transition dynamics has been done for
a strong first-order transition, the experimental
lower bound on  the Higgs mass of 65 GeV \cite{19} has all but
ruled out this regime in the 
context of the minimal standard model. More recently, there has
been interest in the dynamics of very weak first-order
transitions \cite{1}--\cite{6}. In this case the phase
transition may be completed  through the process of phase mixing,
whereby subcritical bubbles effectively restore symmetry,
rather than the conventional nucleation of
critical bubbles.
However,  it is also possible that the dynamics
lie in the intermediate region between a strong transition ---
nucleation 
and expansion of bubbles in a homogeneous background ---
and a weak transition --- phase mixing and domain coarsening.

This work focuses on the dynamics of a transition which lie
in this intermediate region. In this case fluctuations will not
be strong enough to completely dominate the transition and
restore thermal equilibrium, yet may have an effect on processes
such as the nucleation and expansion of bubbles. Though we will
not address the issue of nucleation (see
refs. \cite{5}--\cite{7}  for
discussions on this topic), we will investigate the ramifications of
fluctuations on the speed and structure of the bubble wall
and what the implications are on the standard mechanisms for
generating excess baryons at the electroweak phase
transition. These mechanisms depend strongly on particle
interactions with the bubble wall, and calculating the resulting
baryon abundance requires knowledge of the details of the
transition, including the wall's speed and structure. Some
authors have argued that subcritical bubbles are not relevant to
baryogenesis since fluctuations only have an effect when the
phase transition is only very weakly first order, too weak to be
of interest to baryogenesis \cite{7},\cite{8},\cite{13}. These
arguments, however,  
deal primarily with the question of whether the
phase transition proceeds through the nucleation of critical
bubbles. The issue addressed in here is whether the
fluctuations can have a 
significant effect on critical bubbles after they nucleate.

This issue is related to one which has recently become of interest in
condensed matter physics. Specifically, there has been an effort to
understand how noise affects the properties of front propagation in a
phase transition \cite{19.5}. The results of these studies has varied
depending on the specifics of the model in question. Here, though
fluctuations play the role of the noise in the phase transition, the
system we consider and the model we use to analyze it is significantly
different from those evaluated within the context of condensed matter.

Once one understands the dynamics of the electroweak phase
transition it  becomes possible to calculate the baryon
asymmetry. Traditionally, baryogenesis has been investigated in
two different limits,  thin walls or thick walls. In the thin
wall regime the wall is thin compared to the mean free paths of
the particles, which behave as if they were scattering off a
potential barrier with CP violating reflection coefficients. 
The reflected charge results in a baryon asymmetry in the
region preceding the phase boundary. In the thick wall case the
plasma is treated as though it were in quasistatic thermal
equilibrium. Chemical potentials are introduced for quantities
which vary slower than the time it takes for the bubble wall to
pass, and baryogenesis is the result of a change in the
CP-violating phase, which acts like a chemical potential for
baryon number \cite{15}. In both cases the transition from one vacuum
state to the other is smooth, with the order parameter varying
monotonically. If fluctuations play a significant role it is
likely that there would be deviations away from the smooth
transition model. It is unclear, though, for what transition
strength these deviations can be ignored. Furthermore, our
assumptions about how thick the wall is for a given transition
strength may have to be modified.

The plan of this article is as follows. In Section 2 we discuss
the potential we use in our simulations and its temperature
dependence. Section 3 contains the evaluation of the nucleation
temperature and expansion velocity of critical bubbles in the
standard homogeneous background model in 2+1 dimensions. We
elaborate on our model for the phase transition with thermal
fluctuations in Section 4, covering such issues as the equation
of motion, lattice considerations, and specific
results. Finally, in the conclusion, we discuss ramifications to
electroweak baryogenesis as well as possible improvements for
future work.

\section{The Standard Scenario}
\label{standard}

\subsection{The Potential}
\label{potential}

The potential we select deliberately resembles the temperature
dependent electroweak effective potential for the minimal
standard model,
\be
\tilde V(\phi,T)=\frac{ a}{2}(T^2-T_2^2)\phi^2
  -\frac{\tilde\alpha}{3}T\phi^3+\frac{\tilde\lambda}{4}\phi^4.
\label{effpot}
\ee
The parameters $\tilde\alpha$ and $\tilde\lambda$ determine the strength of
the phase transition and in the minimal standard model are
related to the gauge boson masses and the Higgs
mass respectively. The application of this potential, however,
may be more general and useful in studies of first-order
transitions. 
Due to the 2+1 dimensional nature of the simulations,$\tilde V$ has a
mass dimension of $[M^3]$ with $\tilde\phi$ being
$[M^{1/2}]$. Furthermore,  we
do not associate particular values of $\tilde\alpha$ and $\tilde\lambda$
with particle masses, but rather concentrate on which regions of
parameter space constitute weak and strong transitions as
determined in Ref.\ \cite{2}.

We will find it useful to change from dimensional to  dimensionless
variables as follows, 
\be
x\rightarrow x/ \sqrt{a}T_2
\ee\be
t\rightarrow t /\sqrt{a}T_2
\ee\be
\phi\rightarrow\chi T_2^{1/2}
\ee\be
T\rightarrow\th T_2.
\ee
The potential is now,
\be
V(\chi,\th) = \frac{1}{2}(\th^2-1)\chi^2-\frac{1}{3}\alpha\th\chi^3
      +\frac{1}{4} \lambda\chi^4,
\ee
where $\alpha=\tilde\alpha a^{-1}T_2^{-1/2}$ and
$\lambda=\tilde\lambda a^{-1}T_2^{-1}$. For
$\th<\th_1=(1-\alpha^2/4\lambda)^{-1/2}$ the potential has one
maximum $\chi_-$ and two
minima $\chi_0$, $\chi_+$ located at $\chi_0=0$ and $\chi_\pm=
\alpha\th/2\lambda(1\pm\sqrt{1-4\lambda(1-\th^{-2})/\alpha^2})$.
At the critical temperature
$\th_c=(1-2\alpha^2/9\lambda)^{-1/2}$ the two minima are
degenerate with the minimum at $\chi_+$ being the global minimum
for $\th<\th_c$. At $\th=1$ the minimum at $\chi_0$ disappears.
Figure 1 shows the potential for $\lambda=0.1$, $\alpha=0.4$,
$\th=\th_c - 0.0025=1.2432$.

\subsection{Dynamics}
\label{dynamics}

The dynamics of the phase transition depends to a large degree
on the amount of supercooling undergone before the nucleation of
bubbles. If the supercooling is large, as in the case of a
strong first-order transition, the bubbles expand very rapidly
and have a relatively thick wall. In a weak transition the free
energy of a bubble is minimized for a thin wall and the small
cooling results in a slowly moving wall.
The first step, then, is to determine
the temperature at which nucleation takes place.

As the universe cools below the critical temperature the
symmetric vacuum becomes metastable with a finite probability of
decaying into the asymmetric stable vacuum. The theory of bubble
nucleation from a metastable to a stable state was developed by
Langer \cite{20} and later applied to cosmological phase
transitions \cite{21},\cite{22}. The
equation describing the rate of nucleation per volume in 3+1
dimensions, however, must be changed to account for the
different scaling in
a 2+1-dimensional model. Recall that in
2-dimensions volume scales as $T^{-2}$ and not $T^{-3}$. The
nucleation rate per volume is
\be
\Gamma/{\cal V}=A T^3 e^{-F_c/T},
\ee
where $F_c$ is the free energy of a critical bubble. The rate is
dominated by the exponential; hence, the exact value of $A$ is
not very important and we set it equal to one. The nucleation
temperature is given by the temperature at which the probability
of nucleating a critical bubble inside a horizon approaches
one. We must, therefore, determine the size of the horizon as a function
of temperature. In a 2+1 dimensional universe during the
radiation dominated era, the energy 
density of the universe $\rho$ varies with the scale factor $a$
according to $\rho\propto a^{-3}$. The time-temperature relation
is then
\be
t=\xi \frac{\mpl^{1/2}}{T^{3/2}},
\ee
where $\mpl=1.22\times10^{19}\gev$ is the Planck mass and
$\xi\approx1/30$ \cite{24.5}. The volume inside the horizon at
a temperature $T$ is
\be 
{\cal V}_H \approx 4\xi^2\frac{\mpl}{T^3}.
\ee
{}From this we can write down the probability of nucleating a
bubble inside a causal volume at a temperature $T$,
\be
dP\apr\Gamma{\cal V}_H dt\apr6\xi^2
  \left(\frac{\mpl}{T}\right)^{3/2} e^{-F_c/T}\frac{dT}{T}.
\ee
We define the nucleation
temperature as the temperature for which the total probability
of having nucleated a bubble approaches one,
\be
1\apr4\xi^2 \left(\frac{\mpl}{T_n}\right)^{3/2} e^{-F_c/T_n},
\label{prob}
\ee
where $T_n$ is the nucleation temperature.
Estimating  $T_n$ is made much simpler by
approximating  $T_n$ in the prefactor of the right
hand side of Eq. (\ref{prob}) as the critical temperature.
Because of the exponential, the final answer is not
very sensitive to the value of the critical temperature
chosen. A value 
of $T_c=100\gev$ yields
\be
53\apr F_c/T_n.
\label{53}
\ee

In order to calculate the free energy of a vacuum bubble, we
choose the energy of the meta-stable vacuum as zero,
$V(\vev\phi=0)=0$. The excess free energy of a bubble is
\be
F=\int d^2x\left(\frac{1}{2}(\nabla\phi)^2+V(\phi,T)\right).
\ee
The first term in the integral represents the surface energy of
the bubble, while the second term is the volume energy, coming
from the difference in free energy inside and outside the
bubble. The free energy, in general, must be found
numerically. The difficulty in calculating $F$ arises because
one needs to know the value of the field at all points in space,
which in general may not take a simple functional form. Under
certain limits, however, approximations prove fairly
accurate. In the case of extremely small supercooling the wall
approaches the well known kink solution, the ``thin-wall''
approximation. As the temperature drops further below the
critical temperature the wall shape is reasonably approximated
by a Gaussian \cite{10}. At temperatures appropriate to
nucleation, neither 
approximation turns out to be particularly good. In spite of the
fact that the thin-wall approximation fails to accurately
predict the free energy associated with the nucleation of
critical bubbles, we use it here as a way of estimating a value
for $T_n$. One should bear in mind, however, that the free
energy of the true critical bubble solution is larger than the
free energy in the thin-wall case at a given temperature. As a result
the nucleation 
temperature in the thin-wall case is higher, i.e.,\ less
supercooling, than in the true case. We, however, are not most
interested in what the actual nucleation temperature is, but
rather in the dynamics of bubble expansion. Within this context
the thin-wall approximation gives a reasonable estimate.
Also, recent
studies have shown that the actual nucleation temperature
may be significantly higher than what one obtains by the
standard method \cite{5}, \cite{6}. The basic idea is that in
a weak first-order 
transition large amplitude fluctuations cause the energy
density of the metastable vacuum to shift away from
$V(0)$. Instead one must include a non-perturbative correction
to the free energy of a critical bubble which has the effect of
raising the nucleation temperature. 

To calculate the free energy in the thin-wall limit, when
supercooling is small, we use a perturbative expansion in
$\Delta\equiv1-\th/\th_c$,
the amount of cooling below the critical temperature.
To first order the potential is
\be
V(\chi,\th)=\frac{1}{2}(\th_c^2-1)\chi^2-\frac{1}{3}\alpha\th_c\chi^3
  +\frac{1}{4}\lambda\chi^4
  -\Delta(\th_c^2\chi^2-\frac{1}{3}\alpha\th_c\chi^3).
\ee
A bubble of true vacuum which is just large enough to grow
satisfies the static solution to the equations of motion,
\be
\frac{d^2\chi}{dr^2}+\frac{1}{r}\frac{d\chi}{dr} 
  =\frac{\partial V}{\partial\chi}.  \label{static}
\ee
In the thin-wall limit the spatial derivatives are small except
in the bubble wall. Furthermore, when $\Delta$ is small the
radius of a critical bubble becomes very large and the first
order derivative term in Eq.\ (\ref{static}) becomes negligible. This
gives
\be
\frac{d\chi}{dr}=\sqrt{2V}.
\ee
The free energy of a bubble of radius $R$ is
\be
F\approx\pi R\int_{\chi(R-\delta R)}^{\chi(R+\delta R)}d\chi\, 
  \sqrt{2V} + 2\pi\int_0^Rdr\,r\,V(\chi_+,\th),
\ee
where $\chi_+$ represents the value of the field of the true
vacuum and is given by
\be
\chi_+\approx\frac{2\alpha\theta_c}{3\lambda}
  -\Delta(\frac{2\alpha\th_c}{3\lambda} -\frac{6}{\alpha\th_c}).
\ee
As a function of the radius $R$ and the amount of supercooling
$\Delta$ the free energy is 
\be
F(R,\Delta)\approx \frac{2\pi R}{81}\sqrt{2\lambda}\left(
  \frac{\alpha\th_c}{\lambda}\right)^3 -\frac{4\pi R^2}{9}
  \left(\frac{\alpha\th_c}{\lambda}\right)^2\Delta.
\ee
By taking the derivative with respect to $R$ we get the radius
of a critical bubble
$R_c=\alpha\th_c/(18\sqrt{2\lambda}\Delta)$. Since $\Delta$ is
small, the nucleation temperature is approximately the critical
temperature, and after
equating with Eq.(\ref{53}) we
get for the critical free energy to temperature ratio
\be
\frac{F_c}{\th_n}\approx\frac{\pi}{1458}
  \frac{\alpha^4\th_c^3}{\lambda^3\Delta}
  \approx 53
\ee
or,
\be
\Delta\approx(4.07\times10^{-5})\frac{\alpha^4\th_c^3}{\lambda^3}.
\ee

Once a critical bubble is nucleated it begins to expand because
the free energy lost due to the interior being in the lower energy
vacuum offsets the gain in surface energy. The wall quickly
accelerates, and in the case of a vacuum transition, approaches
the speed of light within a few wall widths \cite{22}. In the more
realistic case where the wall is expanding through a plasma, the
wall experiences an opposing force due to interactions with
particles and reaches a terminal velocity. Calculating 
the terminal velocity proves to be a difficult problem
because it depends on detailed interactions of the Higgs field
with the plasma. Furthermore, the size of the damping also
depends on the thickness of the wall relative to the mean free
paths  of the particles in the plasma
\cite{10}. Simply, this 
results because a thin wall causes particles to make the
transition into the true vacuum state quicker than the time it
takes for them to equilibrate. A thicker wall allows
the particles to maintain quasistatic thermal equilibrium, 
because the thermalization rate is faster than the rate at which the
Higgs field changes,
but not chemical equilibrium, because some particle interaction rates
occur slowly resulting in departures from equilibrium populations. 
The result is a different damping force depending on the regime.
In any case, we assume here that the damping force can be
modeled by a velocity dependent term in the equation of motion
of the Higgs field, where the magnitude of the proportionality
constant determines the terminal velocity,
\be
\dd\chi dd t+\eta\d\chi d t-\nabla^2\chi=-\del V\chi.
\ee
In the frame moving with the wall we can change coordinates to
$\tau=\gamma(x-vt)$, simplifying to the case of a very large
bubble so that we may treat the problem as effectively one
dimensional. The equation of motion then takes the form
\be
\dd\chi dd\tau+\eta\gamma v\d\chi d\tau=\del V\chi.
\ee
The boundary conditions on $\chi$ state that the derivative must
vanish far from the wall. We can then integrate to obtain
\bea
\eta\gamma v\int_{-\infty}^{\infty}\left(\d\chi d\tau\right)^2d\tau
    &=&V(0)-V(\chi_+)\\ 
    &=&-V(\chi_+)
\eea
Within the validity of the thin-wall approximation, we may
perform the integral analytically to obtain 
\be
\frac{1}{6}\eta\gamma v\sqrt{\frac{\lambda}{2}}\chi_+^3 =V(\chi_+).
\ee
Several authors \cite{8},\cite{10},\cite{10.5},\cite{12}
have calculated the velocity of the bubble wall
in both the thick and thin case. The results for a thin wall are
$v\sim 0.1$ and $v\sim0.2$--$0.6$ for a thick wall. Given
a particular set of parameters for the potential, then, we can
find an appropriate value for $\eta$. Here we say that $\eta$ is
generally ${\cal O}(0.1)$.

Up to this point our treatment of the bubble has been from a
purely field theoretic point of view. We describe the dynamics
of the Higgs field as a scalar field in a temperature dependent
potential. Particles in the plasma scatter off the field
effectively creating a damping force. A more macroscopic view of
the phase transition describes the plasma through
hydrodynamics and the bubble wall as an effective combustion
front. In this case there are several effects which arise which
affect the dynamics of the bubble expansion which are not
otherwise evident in the field theoretic point of view. The
velocity of the wall, for instance, depends not only on the
damping coefficient $\eta$, but also on the ability of the fluid
to conduct heat away from the wall and the resulting small
temperature deviations created through the release of latent
heat \cite{10.5},\cite{12}. Furthermore, fluid dynamical
instabilities can arise both 
in the plasma and the bubble wall. Perturbations in the wall may
grow exponentially depending on the size of the perturbation and
the strength of the front \cite{23}--\cite{26}. The net effect
is that bubbles might
not grow in a spherically symmetric way. Nevertheless, including
these effects is beyond the scope of this work and we consider
here only the simplified scenario of a scalar field in a
potential well with damping.

\section{Phase Transitions with Fluctuations}
\label{fluctuations}

Thermal fluctuations play an essential role in the phase
transition, so it is instructive to discuss briefly how they
enter into the physical picture. Without fluctuations the field
remains trapped in the metastable vacuum until the potential
barrier vanishes at which time the field rolls down the
potential into the true vacuum. In a thermal environment, 
though, it is only the expectation value of the field which is in
the metastable state. There is a finite probability that the
field can take on a value beyond the potential barrier and thus
make the transition into the true vacuum. Hence, the nucleation
and expansion of bubbles. This method, however, traces the
evolution of the field using the equations of motion
at zero temperature often with a damping term to simulate the
dissipation due to the thermal environment, with finite temperature
effects limited to corrections to the effective
potential. This approach is justifiable if one wishes to 
consider only the behavior of the expectation value of the field,
where one has implicitly assumed that the variance of the field
value is very small. This assumption should hold if the minima
in the effective potential are sufficiently deep that thermal
fluctuations away from the minima are strongly suppressed. Even
in this case, however, the dynamics of the field may not be
adequately described by the equations of motion. For instance,
if the field is initially located at a local minimum of the
potential, which then becomes a local maximum in such a way
that the slope remains zero, the field remains in this unstable
extremum indefinitely. Realistically, what happens is that
fluctuations dislodge the field from the extremum, which
subsequently evolves according to the equations of motion.

The question of how one is to model the thermal fluctuations in
field theory nevertheless remains. In classical statistical
mechanics one treats the problem of thermal fluctuations
by identifying a system
and a heat bath and choosing a model for the coupling between
the two and solving for the behavior of the system.
Finding a solution is greatly simplified when the reservoir obeys
particular criteria. Specifically, if
the reservoir has infinite specific heat and a relaxation time
much shorter than that of the system, it
remains in thermal equilibrium even though it interacts with
a system which may not be.
When these conditions are satisfied one may ignore the dynamics
of the reservoir in favor of the dynamics of the system. This
paradigm provides a method for calculating quantities such as
equilibration time scales of a system and transition probabilities
when the system is in equilibrium. In the case of
Brownian motion, for example, where
the system is
a small macroscopic particle and the bath  is the fluid in
which it rests,
the Langevin equation is a
natural outcome of the fluctuating thermal force of the bath on
the  system. 

In the case of field theory, however, the dichotomy of the
system and bath is less clear. If there is only one
self-interacting field one can decompose the dynamics into short
and long-wavelength modes which operate on different
time scales. The short wavelength modes respond much more
quickly and can serve as the thermal bath while we define the system
as those modes whose wavelengths are larger than some critical value;
non-linear interactions
couple the system to the bath. Another approach is to have a
second field which acts as the bath to the first field. In
either case the form of the coupling determines the nature of the
fluctuations which the system experiences. 
Below, we model the fluctuations as white noise (uncorrelated)
by adding a stochastic term to the equation of motion. In
general, the noise which the system experiences may be
significantly more complicated, as is the case in some simple
cases which have been studied \cite{28},\cite{29}. We justify
our choice by noting that it is not yet clear how these findings
might change given the relatively complicated environment of the
early universe. Furthermore, in at least on study \cite{29}
the authors found that in the high temperature limit the noise
becomes white.

The Langevin equation is a popular way to model phenomenologically
a system with thermal fluctuations, though, as mentioned
above, the actual dynamics of the noise may be different from
the simple model of additive white noise. Nevertheless, the
crucial question is deciding how one is going to model the essential
physics of the system. Certainly, if we knew what the true
equations of motion for the Higgs and particle fields were in the early
universe,  we could use those
equations. This, however, is not the case. At the time of the
electroweak phase transition, aside from the Higgs, there were
many other particles, and writing the  complete dynamics is,
if not impossible, extremely difficult. Given this situation, using the
Langevin equation with additive white noise is a reasonable
place to start in the investigation of how fluctuations may
affect the dynamics of the transition, and we use it here.
Also, numerical studies of nucleation of critical bubbles in 1+1 and
2+1 dimensions carried out using the Langevin equation
to model thermal fluctuations have demonstrated good agreement
with classical nucleation theory \cite{30},\cite{31}.

\subsection{The Model}

As described above the coupling of the field with the thermal
bath is modeled by a Langevin equation. The equation of motion,
then, in terms of the dimensionless variables is
\be
\ddel\chi t=\nabla^2\chi-\eta\del\chi t-\del V\chi+\xi({\bf x},t).
\ee
where we have defined $\xi$ as the dimensionless stochastic
noise. The fluctuation-dissipation theorem relates the noise to
the dissipation coefficient $\eta$ by
\be
\langle\xi({\bf x},t)\xi({\bf x}',t')\rangle
=2\eta\th\delta(t-t')\delta({\bf x}-{\bf x}').
\ee
In discrete form, $\xi$ is
\bea
\xi({\bf x}_i,t_n) &=& \xi_{i,n}\\
  &=& \left(\frac{2\eta\th}{\delta t(\delta x)^2}\right)^{1/2}{\cal G}_{i,n} 
\eea
where ${\cal G}_{i,n}$ is a unit variance Gaussian random number
at each point on the lattice and $\delta t$ and $\delta x$ are the
lattice spacing in the time and spatial directions, respectively. We
integrate the equation of 
motion  forward in time using a second-order leap-frog method.

We carry out the numerical experiment by inputting initial
conditions and allowing the simulation to run. Since we are
interested in the dynamics of the bubble wall and not 
questions of nucleation, we use as initial conditions a wall
which stretches across the width of the lattice located at the
midway point along its length. Half of the initial lattice
volume, then, is in the asymmetric phase while half is in the
symmetric phase. The shape of the wall is chosen so that it
conforms to the kink profile appropriate for a thin wall. This
profile is in fact not the one which minimizes the free energy,
but it is sufficiently close so that the time it takes for the
wall to relax to its correct form is much smaller than the run
time. By making the wall stretch across the width of the lattice
we effectively model a large bubble and thus ignore the initial
expansion stage following nucleation. The wall, however, is not
given any initial speed, but must accelerate to its terminal
velocity. The time for this to happen is also very small.

The initial configuration of the wall is prepared without noise
added in. We include the fluctuations in the simulation only
through the dynamics of the equation of motion. Though such a
configuration is certainly unphysical given the assumption that
thermal fluctuations exist, it has no  effect on the
outcome of the numerical experiment. In each of the two phases
there exists a thermal equilibrium distribution of the other
phase due to fluctuations \cite{2},\cite{3}. Given the
strengths of the transitions we consider, the time it takes to
reach this distribution is small. Only during the initial stages
of the simulation is the bubble wall likely to be affected by
particular characteristics of the initial conditions.

A first-order phase transition may be classified into weak and
strong transitions depending on the height of the barrier
separating the minima. In a weak transition considerable phase
mixing may exist with the transitions proceeding through domain
coarsening, while in a very strong transition we expect the theory of
homogeneous nucleation and bubble expansion to be correct. We
wish to explore here the intermediate region and thus must
choose appropriate values for the parameters which describe the
potential. Based on the findings in Ref.\ \cite{2} we select
values for the 
parameters $\alpha$ and $\lambda$ which explore transitions in
the strong regime, where a strong transition is one where 
at $\th_c$ the symmetric phase comprises at least about 60\% of the
total area. We, therefore, set
$\lambda=0.1$ and allow $\alpha$ to take on the values
0.4 and 0.5, well into the strong regime.

The potential in Eq.\ (\ref{effpot}) has a direct connection with the
electroweak phase transition. In the temperature one-loop
effective potential $\alpha$ is related to the gauge-boson
masses while $\lambda$ is determined by the Higgs mass. The
values we choose here for these parameters, however, should not
be construed as exploration of the parameter space of masses of
these particles. The naive interpretation that the chosen values
of $\alpha$ and $\lambda$ correspond to masses  in the minimal
standard model is false in this case because the simulations
describe dynamics in a 2+1 dimensional world. What constitutes a
weak and strong transition in 2+1 dimensions is different from
the 3+1 dimensional case. We choose, therefore, not to make any
claims about particle masses, but rather focus qualitatively on
the effects of fluctuations on the phase transition.

\subsection{The Lattice}

When taking a numerical approach to this problem one must make a
choice regarding the coarse-graining scale as realized
through the lattice spacing. Modes with wavelengths shorter than
the lattice spacing couple to the larger wavelength modes
through the noise term in the equation of motion. The results
from placing the field theory on a lattice, then, only apply to
the long wavelength modes. When probing physics at shorter
wavelengths, one must be careful in taking the continuum
limit. To do so one must  include renormalization
counterterms in  order to obtain the proper continuum
theory. These issues are discussed in more detail in
Ref.\ \cite{30}. Following the prescription in Ref.\ \cite{2} we
set the lattice 
spacing $\delta x = 1$, which is approximately equal to the
mean-field correlation length given by
$l_{cor}^{-2}=V''(\chi_0,\th_c)$. 

In addition to choosing a grid size on the lattice, one must
choose the size of a time step, the size of the lattice, and the
boundary conditions. Ideally a large time step would allow one
to integrate the equations of motion for a longer period of
time; however, stability considerations limit the size of
$\delta t$. To find an appropriately sized step we compared simulation
results as a function of different values of $\delta t$, choosing a
value in the regime where the results
become independent of the magnitude of step. In all simulations
we used a value of $\delta t=0.2$

As with the choice for $\delta t$, selecting a lattice size is
another exercise in compromises. The physics one is attempting
to simulate takes place in an effectively infinite volume, but
one is limited to not just finite lattices but also ones
which are fairly small because of the need for reasonable
integration times. There are dangers, however, in having too
small a lattice. Strictly speaking, in the context of phase
transitions symmetry breaking only occurs in the infinite
volume limit. Within any finite volume there exists the
possibility that a fluctuation will restore the symmetry
regardless of the dynamics internal to the volume. Since it is
precisely these dynamics with which we concern ourselves, it is
paramount to choose a volume large enough that the probability
of such an occurrence becomes negligible. Essentially, what
happens is that fluctuations in the broken phase may drive the
system back to the symmetric phase, even though it is
energetically unfavorable. If the total volume is large, such
fluctuations result in only a small region of the total volume
having its symmetry restored. The chance that this could happen
with a large volume is exponentially suppressed because of the
large amount of energy necessary. We can estimate how large a
volume we need by calculating the rate for a large fluctuation,
\be
\Gamma\sim e^{-F/\th}.
\ee
The free energy $F$ is given by the change in effective
potential energy between the two minima multiplied by the amount
of volume in the asymmetric phase. Ignoring the gradient portion
of the free energy, which only increases $F$ and makes a
symmetry restoring fluctuation less probable, we have
\be
\frac{F}{\th}\approx{\cal V}
  \frac{4\alpha^2\th_c\Delta}{9\lambda^2}.
\ee
where ${\cal V}$ is the volume in the symmetric phase. The
probability of having  a large fluctuation given the size
of our lattice and run time is approximately $10^{-10}$.

Since we are interested in the dynamics of the bubble wall as it
expands, we also want to make sure that it is  unlikely that
a critical bubble will nucleate in the asymmetric phase within
the lattice during a run time. Because the volume under
consideration is much less that a horizon volume and the time is
much less than a Hubble time, the probability of this happening
is similarly suppressed. The lattice size we use in the
simulations is $L_x=100,\ L_y=50$, where $L_x$ is the length in
the $x$ direction and $L_y$ is the length in the $y$ direction.

Another consideration lies in the fact that we are modeling an
unbounded system by a finite volume with a boundary. Though, as
discussed above, we do not expect finite size effects to be
important, there is a distinct surface in the simulations which
is unphysical. We circumvent this issue by using periodic
boundary conditions in the $y$ direction and `open' boundary
conditions in the $x$ direction. The periodic conditions allow
us to  model an essentially planar wall, appropriate for a large
bubble. The danger, however, with periodic conditions is that
long range correlations may be induced if the simulation is run
for longer than $L_y/2$, the light crossing time. It turns out that
this is not a concern 
here because the presence of dissipation and noise
have the effect of damping out and swamping any long
range `communication' which might otherwise exist. The open
boundary conditions consist of assuming that for points
immediately outside the lattice the field takes on a value equal
to the field on the boundary. Though not realistic, these
boundary conditions provide an approximation to the unbounded
system. Any spurious effects caused by these conditions do not
extend into the lattice because of dissipation and noise.

\section{Results}
\label{results}

The focus of the numerical experiment is to understand the
effect of the fluctuations on the rate at which the old phase
is converted into the new phase and how that transition is made. 
We expect that for a relatively stronger transition the
fluctuations will play a fairly minor role, while for a
relatively weak transition there may be a marked difference from
the homogeneous background case. We investigate the rate of phase
conversion for different transition strengths by allowing
$\alpha$ to vary while holding all other parameters constant and
comparing the results to the case where there are no
fluctuations. The values of $\alpha$ we choose (0.4 and 0.5) place
the transition  in the strong regime as defined in Ref.\
\cite{2}. In that study the author found that the change from
a weak to a strong transition is itself a second-order phase
transition with a critical value of $\alpha_c=0.36$. We can
characterize the strength of the transition by $f_+$, the
fraction of volume which fluctuates from the symmetric phase beyond the
maximum of the potential barrier. At the critical value
$\alpha_c$ this fraction is 42\%, while for $\alpha=0.4$,
$f_+\approx 6\%$ and for $\alpha=0.5$, $f_+\approx0.1\%$.
Furthermore, whereas in the homogeneous background case the
field in a region of space smoothly makes the transition from
the symmetric to asymmetric phase, we do not expect this to
happen when the amplitude of the noise becomes large.

In order to calculate the rate of phase transformation for a
particular transition we first introduce a definition which will
allow us to determine which phase the field is in. We label the
field at a particular lattice point as being in either the 0-phase if
$\chi\leq\chi_-$ (i.e.\ to the left of the maximum) or the
+ phase if $\chi>\chi_-$ (i.e. to the right of the maximum).
This allows us to determine what fraction
of the total volume is in each of the two phases. Because we
place the bubble wall through the center of the lattice
approximately half of the initial volume is in each phase with
deviations away from half due to fluctuations. Lattice points
move on average from the 0-phase to the + phase as the
bubble expands. In Figures 2--3 we plot the position of
the wall as a function of time for a phase transition with and without
fluctuations for $\alpha=0.4$ and $\alpha=0.5$ with
$\eta=0.2$. The wall position for the case with fluctuations as
shown by the short-dashed line
is an ensemble average of 200 separate trials while the
long-dashed lines show the one standard deviation width of the
distribution. The dotted line shows one realization. The solid
line is the position for a phase transition without
fluctuations, where the step-like nature is due to the
discreteness of the lattice. 
For the case with fluctuations 
the position does not mean that the
midpoint of the wall 
is located at that particular $x$-position all the way across
the lattice rather it represents an average position at that
point in time. In fact there are regions both in front of and
behind the wall which belong to the opposite phase resulting in
a somewhat amorphous boundary. Figs. 4a--d show contour plots of
the field on the lattice for $\alpha=0.4,$ $0.5$ with
$\eta=0.1$, $0.5$ with contours at
$\chi_-/2$ (dotted), $\chi_-$ (solid), $3\chi_-/2$
(dashed). We see significant distortions in the wall away from
planar in the weaker transitions whereas the stronger transitions
approach the plane solution, though other structure is still evident.
The rate at which
phase is converted from the 0 to the + state is also clearly
elevated relative to the 
transformation without fluctuations. 
In Table 1 we show the rate of phase conversion, essentially the
velocity of the wall, for the different transition
parameters. The right-hand columns for a given value of $\alpha$ and
$\eta$  show the
average wall velocity for the fluctuating background case along
with the one sigma width; the left-hand columns show the wall velocity
for the homogeneous background case.
In the limit of a very strong transition we can see
that the rate of phase conversion approaches the homogeneous
theory solution.

Also particularly relevant to electroweak baryogenesis
is the actual structure of the wall. Whether and how many
baryons are created depends on how particles, which are
interacting with the changing Higgs field, make the transition
from the 0 to the + state. Though this depends on quantities
such as the velocity of the particle relative to the wall, which
we don't calculate here, we can still get an idea of what a
particle ``sees'' as it moves from one state to the other by
taking cross sections of the wall. In Figs.\ 5--6 we 
plot the values of the field $\chi$ as a function of $x$ at a
particular time
for $\alpha=0.4$ and $\alpha=0.5$ with $\eta=0.1$ for both the
homogeneous  and fluctuating background theories. While the wall
thickness in the homogeneous theory can be clearly defined, it
is less apparent in the fluctuating case.
We see that averaged over 200 simulations the wall is thicker
relative to the case where there are no fluctuations, though the
increase is only moderate for the transition strengths used here
and approaches the homogeneous background solution for a
stronger transition.

Figures 7--8
follow the behavior of a lattice point as a function of time.
The boundaries of when the transition from the 0 to + state begin
and end is less clear in the fluctuating case as compared to the
homogeneous case. The results from averaging over many
simulations indicates that in general a lattice point makes the
transition more slowly than if there were no fluctuations. This
results because in the vicinity of the
bubble wall a lattice point may undergo several ``transitions''
before finally reaching equilibrium in the + phase.

\section{Conclusions}
\label{conclusions}

We see that quantities such as the rate of phase conversion and the
structure of the bubble wall can be  different for
phase transitions with fluctuations as compared to the
standard homogeneous background model. The speed of the bubble
wall may be increased by a factor of about two for a ``mildly'' strong
transition and probably even more for  weaker transitions.
Physically, this is a reasonable expectation; the wall
effectively ``swallows'' the field fluctuations, moving forward
as it does so. In a transition with more large amplitude fluctuations
this swallowing effect is more
pronounced. Though the rate at which phase is converted from the
0 to the + phase is increased globally, locally the transition
is not necessarily well defined and more gradual.
Furthermore, we see that this effect is noticeable
even though only a small percentage (about $6\%$) of the symmetric
phase may fluctuate beyond the barrier at any given time. 
As the
amplitude of the fluctuations decrease the bubble behaves as
predicted by the homogeneous background theory. That there are large
amplitude fluctuations which can roughly ``mask'' the transition
is not surprising; in fact, this is 
entirely determined by the stochastic noise term in the equation
of motion. What is interesting is that the noise should have
such a significant effect on a transition which one would
consider ``strong'' by the definition given in
Ref. \cite{2}.

It is not necessarily evident that random fluctuations should result
in an increase in the velocity of the bubble wall. Though fluctuations
from the 0 to + phase might speed up the wall, one might also expect
that fluctuations from the + to 0 phase would have the opposite
effect, resulting in no net change. It is the asymmetry in the
potential, however, which prevents this from happening. Because
fluctuations are a result of a random impulse on the field at a
particular point, a fluctuation which drives the field from the 0 to +
phase is more likely to approach or exceed the maximum of the
potential than those that go from the + to 0 phase. The dominant
effect on the dynamics, then, is to enhance to 0 to + transition.

The implications for baryogenesis stem from the fact that
models rely on particular
assumptions of the bubble wall structure, e.g., a
smoothly varying monotonic order parameter. The standard picture
\cite{13} states that, depending on
the thickness of the wall, one  selects a mechanism with which
to calculate the baryon number generated. In order
for this to work, however, it is necessary that the homogeneous
background theory of bubble nucleation and expansion be valid.
Previous studies have called into question this paradigm by
pointing out that subcritical bubbles may affect the initial
conditions of the transition as well as bubble nucleation
\cite{1}--\cite{6} in a weak first-order phase
transition. Here we investigate transitions which are stronger
than those in \cite{2}, but still within the regime where
nucleation may be affected by the presence of subcritical
bubbles \cite{6}.
Although the electroweak transition is most likely
weak in the minimal standard model, authors \cite{7}, \cite{8},
\cite{13} have argued that baryogenesis is likely only for a
sufficiently strong transition where the effects of phase mixing
due to thermal fluctuations would be negligible. 
What we have shown here is that the realm of the phase
transition for which fluctuations may play a significant role is
larger than what was expected. Their effect is not limited to
possible alterations in the picture of homogeneous nucleation or
degree of phase mixing,
but also includes the dynamics subsequent to nucleation. For
instance, not only are the wall thickness
and velocity larger, but the path from the symmetric phase to
the asymmetric phase is hardly smooth or monotonic. Any model
for electroweak baryogenesis will 
clearly have to take the stochastic nature of the dynamics into
account when
considering the interaction of particles with the Higgs
field. The results here are not meant to be quantitative, but
they do demonstrate the importance of fluctuations in the
dynamics of first-order phase transitions.

Finally, we point out some of the limitations of this
work. Ultimately, we would like to be able to obtain a complete
picture of the electroweak phase transition in order to
determine the viability of generating the baryon asymmetry of
the universe at this scale. What we have done here is to examine
the role thermal fluctuations are likely to have on the transition
dynamics, and what the implications are on models of
baryogenesis. However, this study was limited to 2+1 dimensions
rather than the 3+1 dimensions of the real world. Extending this
work to higher dimensions would be a valuable contribution to
our understanding of this problem. Also, we have assumed here
that one can model the thermal fluctuations through stochastic
white noise, as one does when studying phenomena such as
Brownian motion. Realistically, this is only an approximation of
more complicated couplings between system and bath. Indeed, how
one is to divide the physical environment into system and bath
and the nature of the noise that results
is a problem currently under active investigation \cite{28},
\cite{29}. Lastly, the dynamics of the cosmological fluid have
been omitted. Not only should the fluctuations have an effect on
the fluid, but, as already noted, the fluid itself plays a role
in the overall dynamics. A more complete treatment would include
fluid velocity, pressure and temperature as well as parameters
for heat capacity and conductivity in the simulations.
In spite of these caveats, we believe that our
findings are  indicative of what we may eventually
find to be the ``true'' dynamics.

I would like to thank Edward Kolb, Mike Turner, Ed Blucher and Emil
Martinec for their valuable input and advice and Marcelo Gleiser for
his helpful comments. M. A. was supported in part by the DOE at
Chicago and by NASA grant NAG 5-2788 at Fermi National
Laboratory. This work was carried out in partial fulfillment of the
requirements for a Ph.D. at the University of Chicago Physics
Department.

\pagebreak

{\Large \bf Figure Captions}
\begin{description}

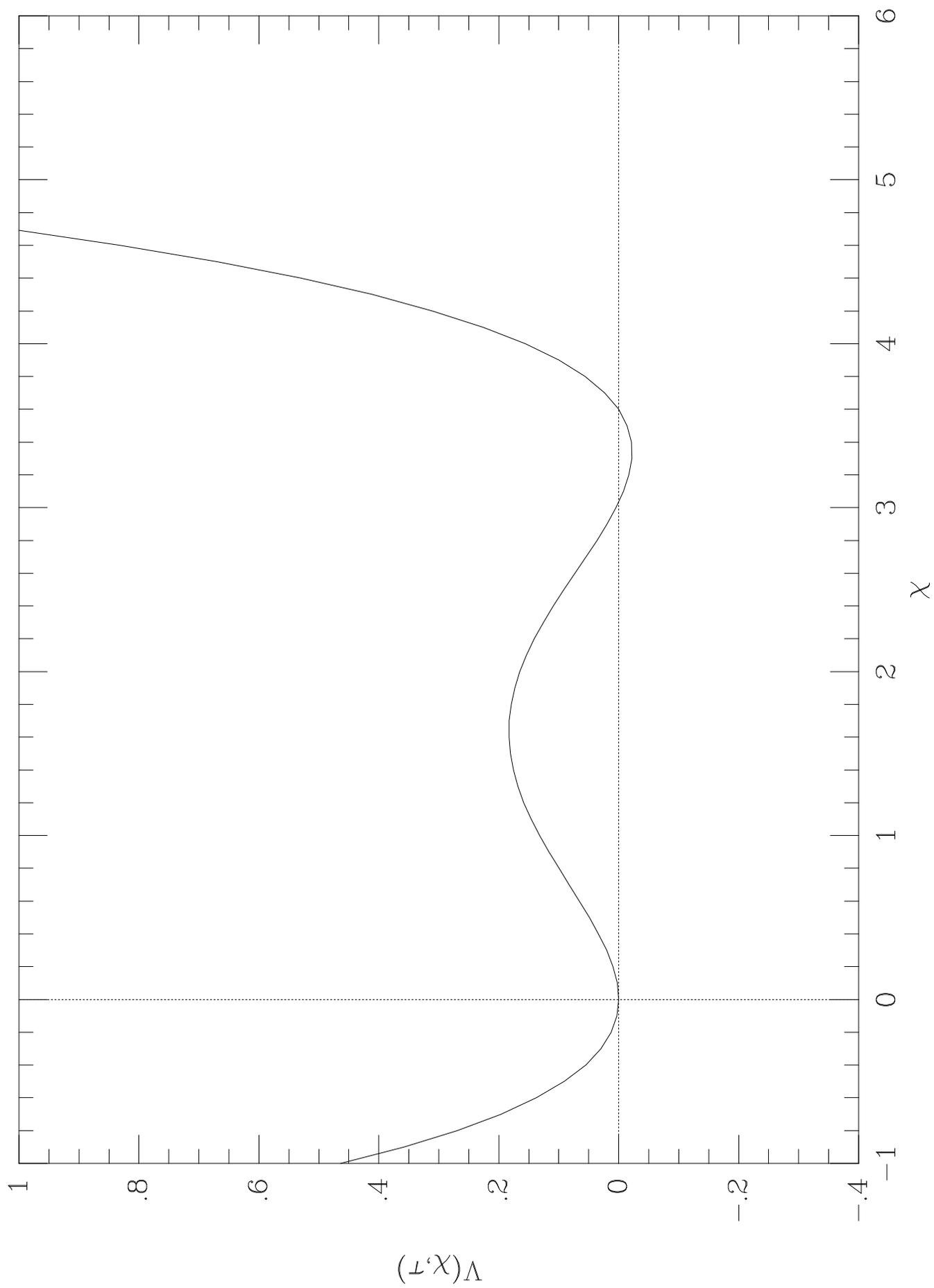
\item[Figure 1.] Potential for $\lambda=0.1$, $\alpha=0.4$,
$\th=\th_c-0.0025 = 1.2432$.

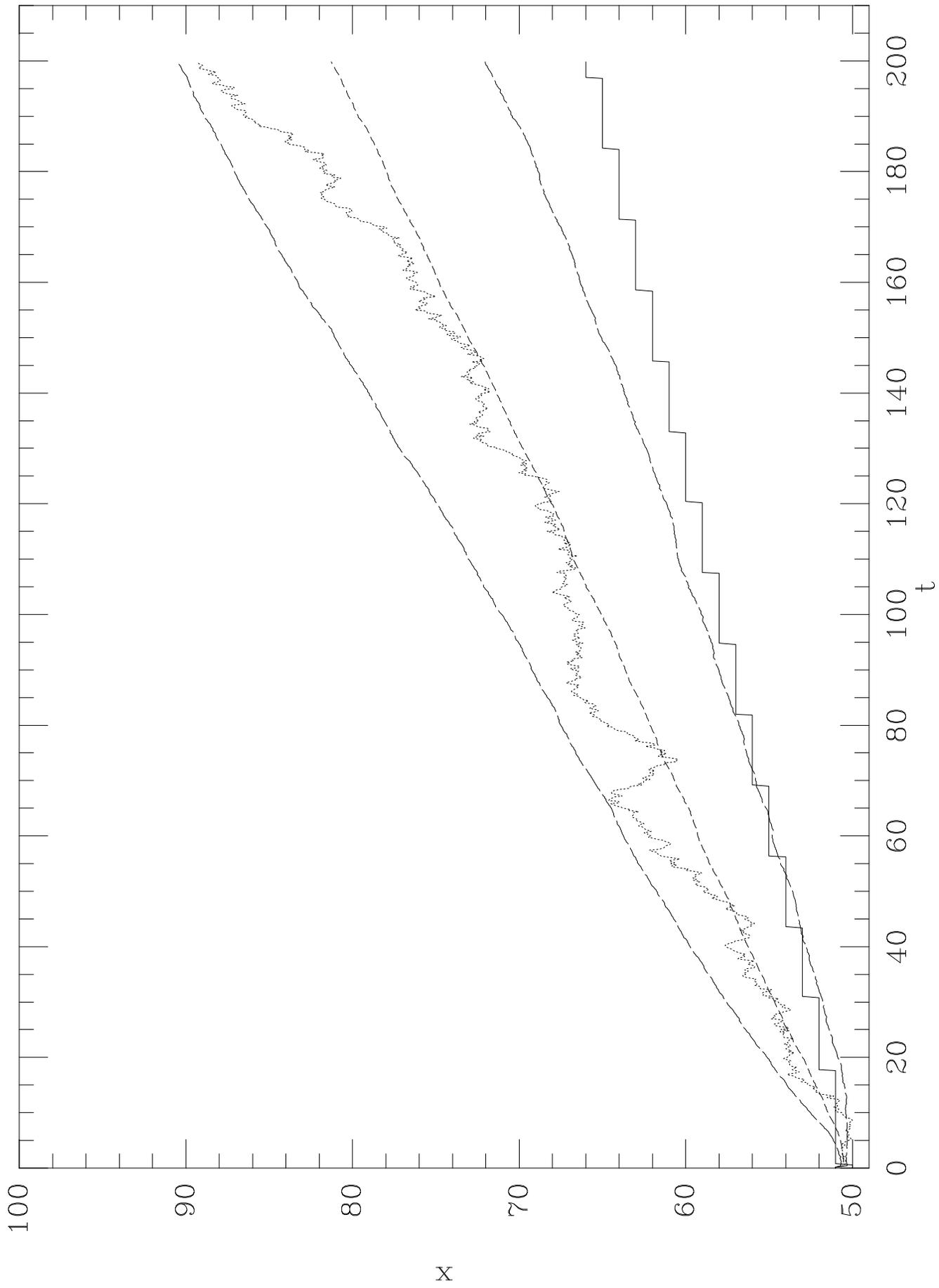
\item[Figure 2.] Average wall position as a function of time for
$\alpha=0.4$, $\eta=0.2$. Shown are a phase transition without
fluctuations (solid line), with fluctuations (dotted line), average
over 200 simulations (short-dashed line), and one-sigma width
(long-dashed line).

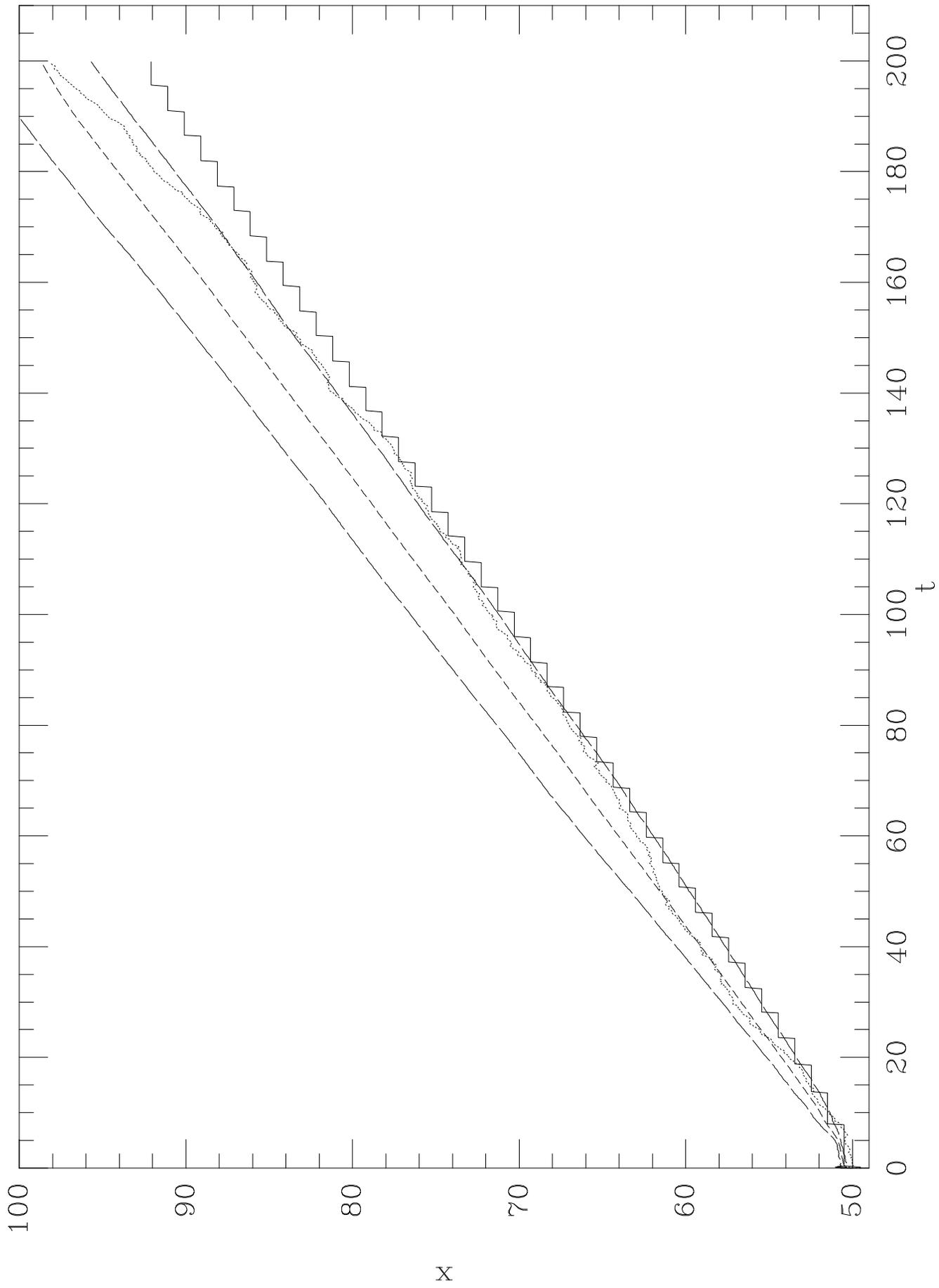
\item[Figure 3.] As in Fig. 2 with $\alpha=0.5$.

\item[Figure 4.] Contour plot of field $\chi$ on the lattice with
contours $\chi_-/2$ (dotted), $\chi_-$ (solid), $3\chi_-/2$ (dashed).
a) $\alpha=0.4$, $\eta=0.1$. b) $\alpha=0.4$, $\eta=0.5$. c)
$\alpha=0.5$, $\eta=0.1$. d) $\alpha=0.5$, $\eta=0.1$.

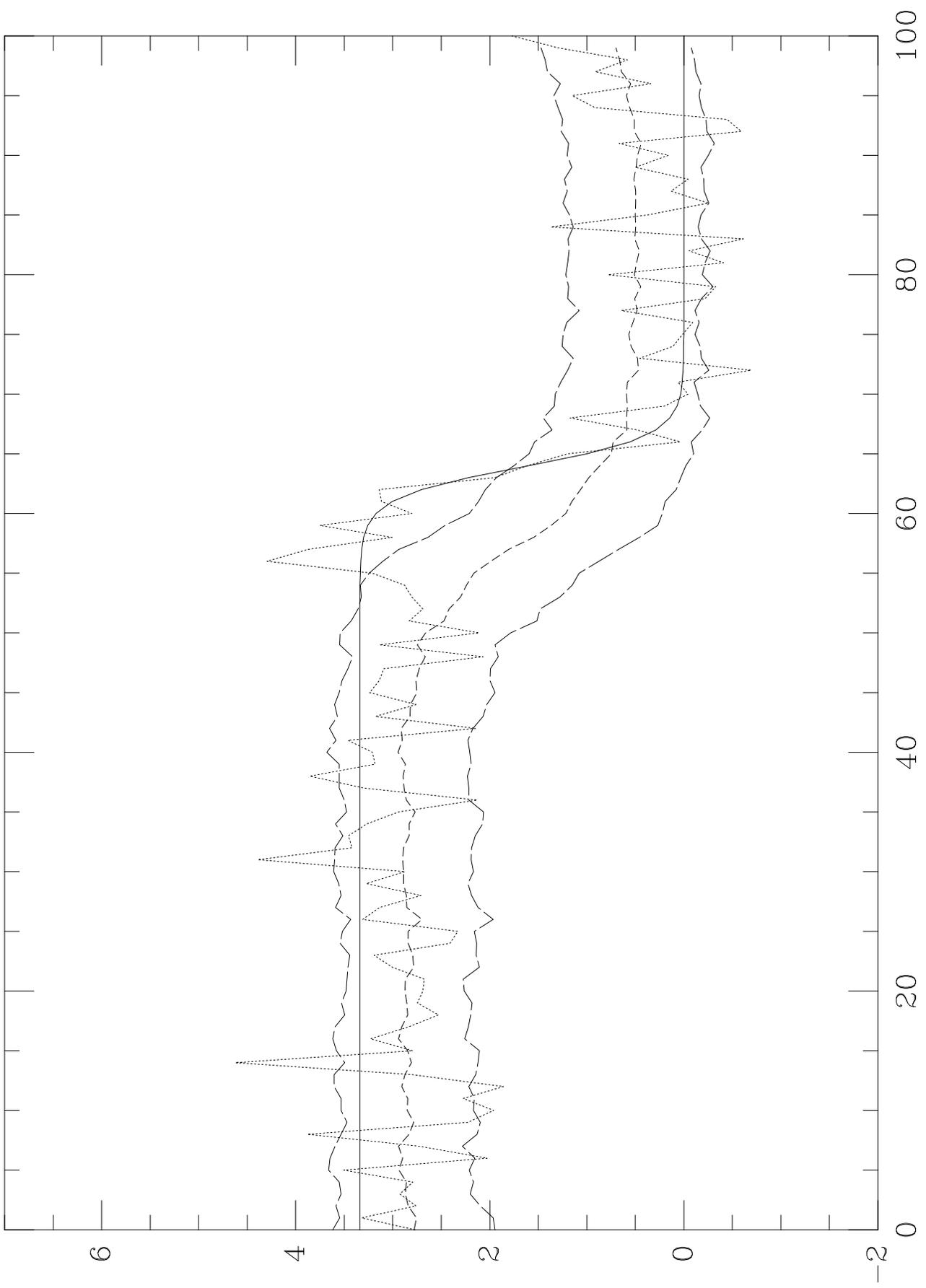
\item[Figure 5.] Wall profile for $\alpha=0.4$, $\eta=0.1$. Shown are
simulation resluts for a phase transition without
fluctuations (solid line), with fluctuations (dotted line), average
over 200 simulations (short-dashed line), and one-sigma width
(long-dashed line).

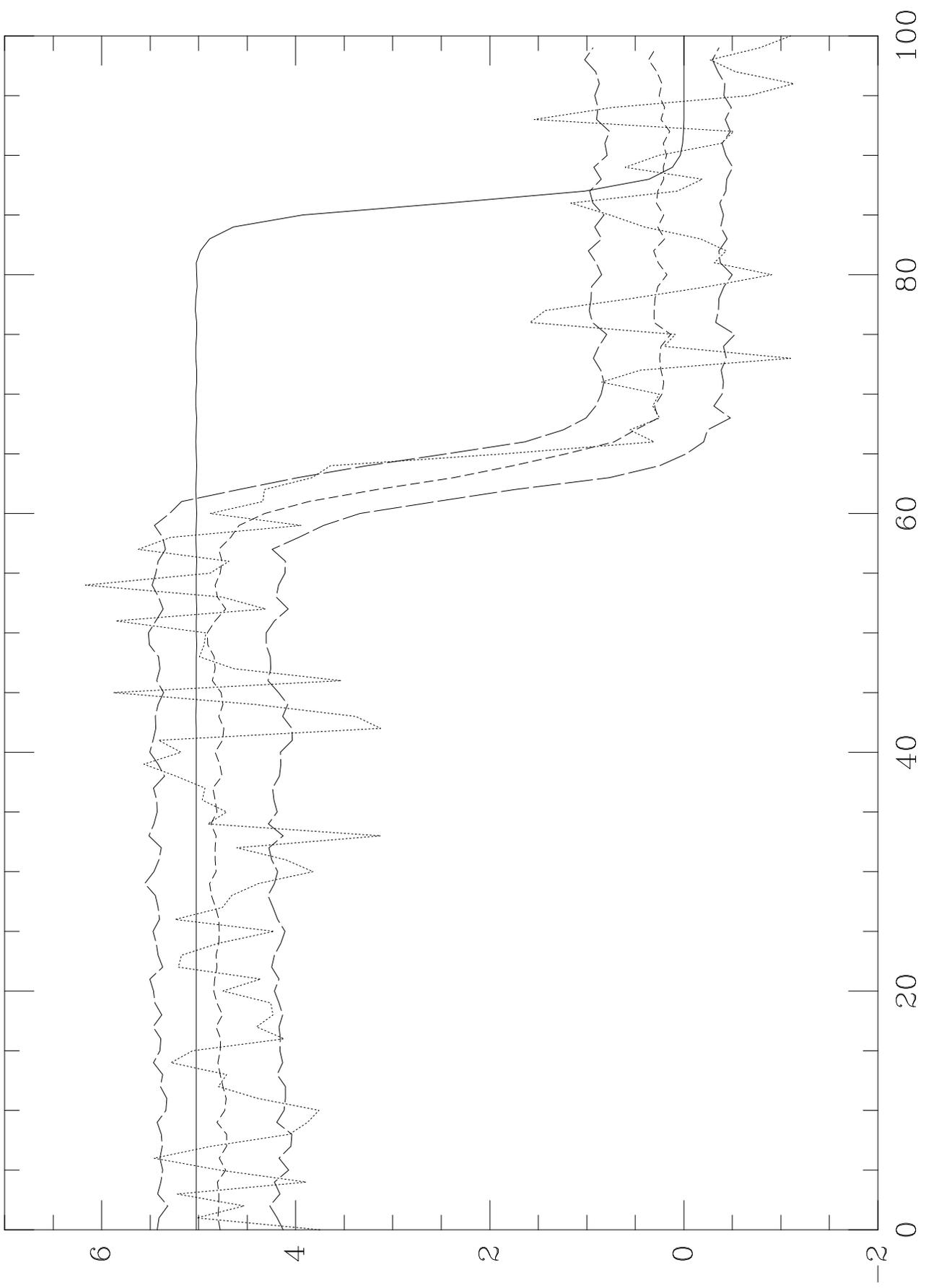
\item[Figure 6.] As in Fig. 5 with $\alpha=0.5$.

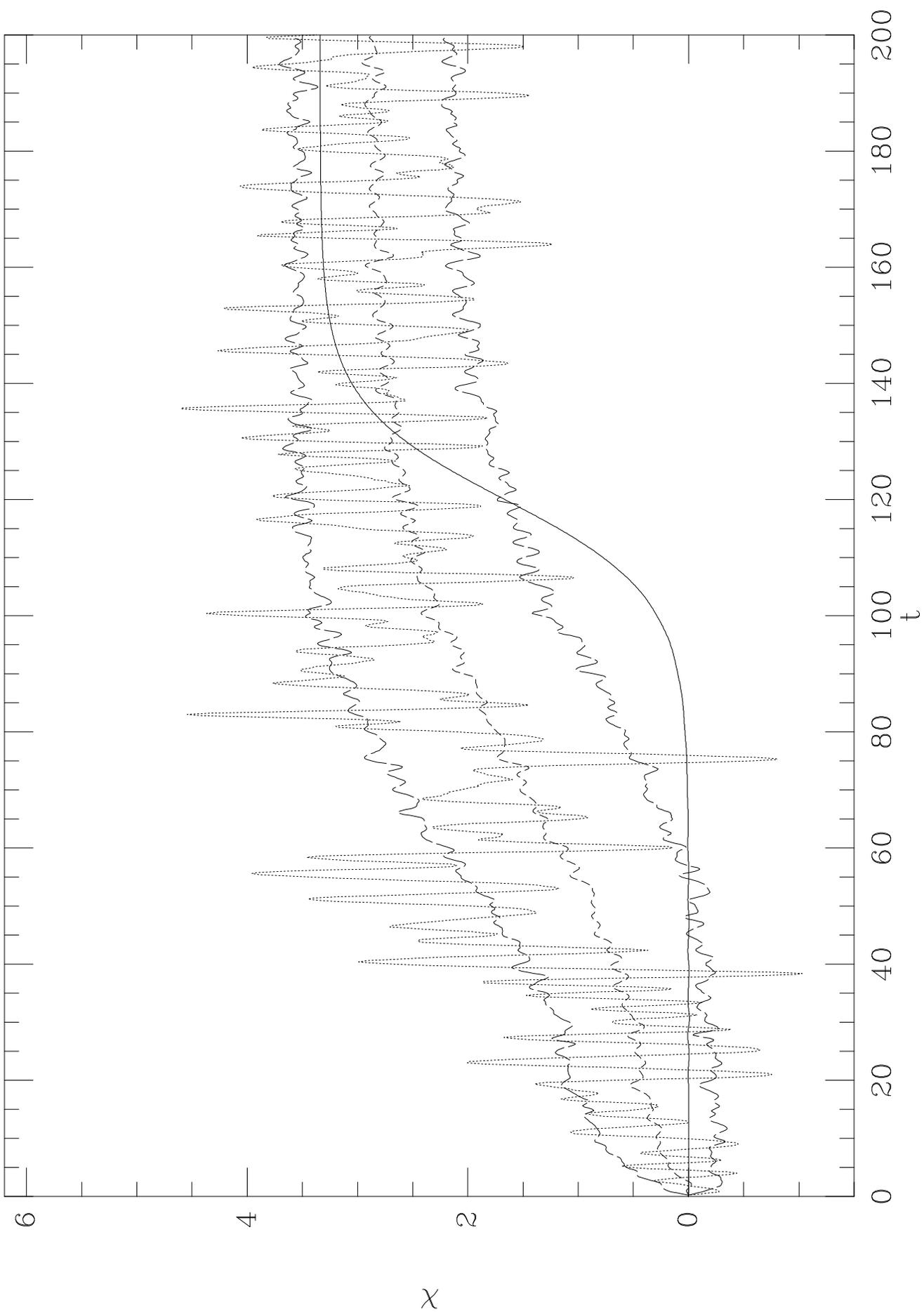
\item[Figure 7.] Lattice point history for $\alpha=0.4$,
$\eta=0.1$. Shown are simulation results for a phase transition without
fluctuations (solid line), with fluctuations (dotted line), average
over 200 simulations (short-dashed line), and one-sigma width
(long-dashed line).

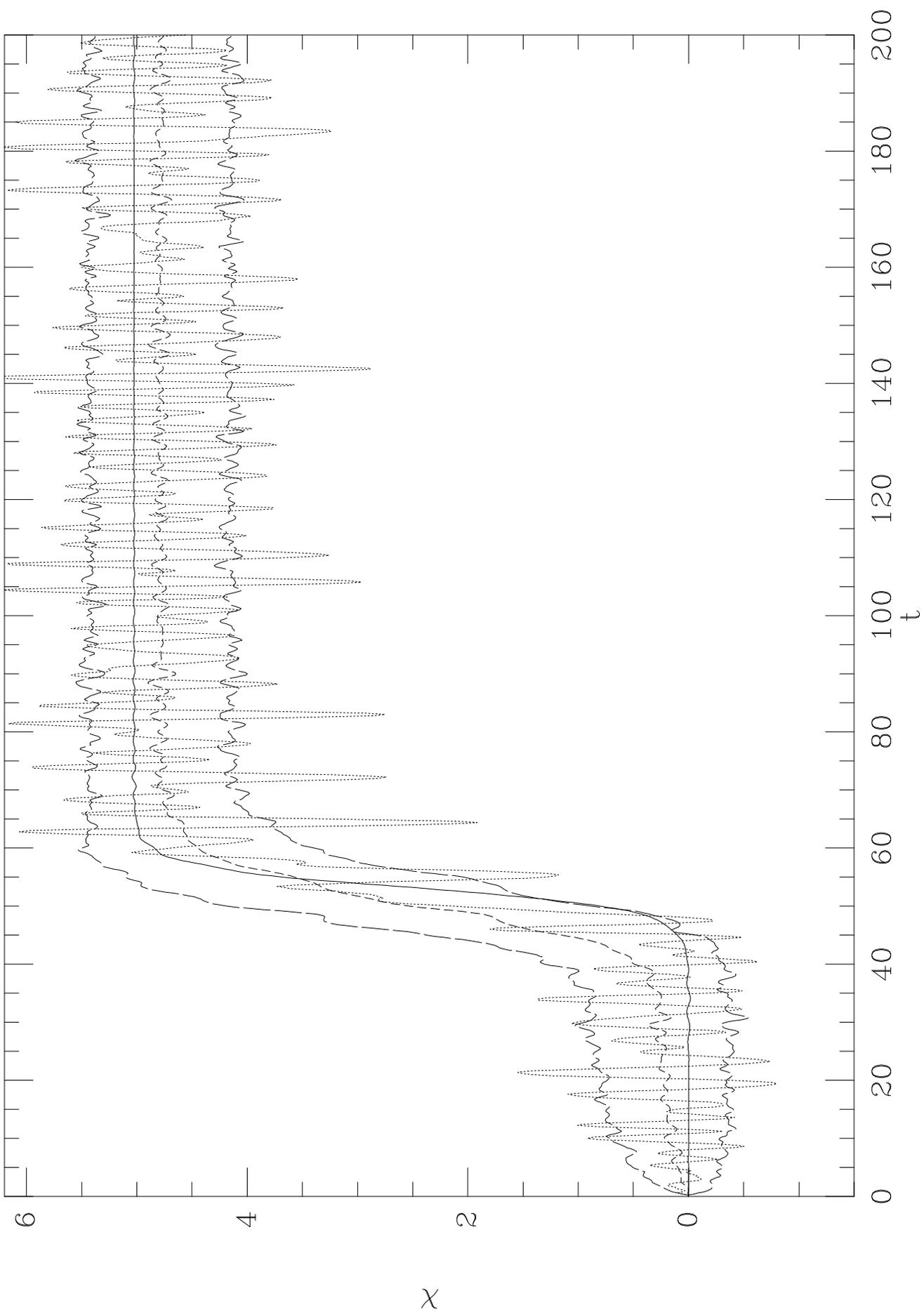
\item[Figure 8.] As in Fig. 7 with $\alpha=0.5$.

\end{description}

\pagebreak

{\Large \bf Table Caption}

\noindent
Table 1. Wall velocities for varying parameter values where $v_h$
represents the velocity in simulations without fluctuations and $v_f$
represents the velocity in simulations with fluctuations.

\pagebreak

\begin{center}
\begin{tabular}{|c||c|c||c|c||c|c|} \hline
  & \multicolumn{2}{c||}{$\eta=0.1$} &
    \multicolumn{2}{c||}{$\eta=0.2$} &
      \multicolumn{2}{c|}{$\eta=0.5$} \\ \hline
  $\alpha$  &  $v_h$ & $v_f$ & $v_h$ & $v_f$ & $v_h$ & $v_f$ \\ \hline
 0.4  & 0.15 & 0.24$\pm$0.06 & 0.08 & 0.16$\pm$0.05 & 0.035 &
    0.077$\pm$0.036 \\
 0.5 & 0.4 & 0.40$\pm$0.06 & 0.21 & 0.25$\pm$0.02 & 0.09 &
    0.12$\pm$0.01 \\ \hline
\end{tabular}
\end{center}

%

\begin{thebibliography}{99}
\addcontentsline{toc}{section}{References}

\bibitem{1} M. Gleiser and E.W. Kolb, Phys. Rev. D {\bf 69},
1304 (1992); G.~Gelmini and M.~Gleiser, Nucl. Phys. {\bf B 419},
129 (1994); M.~Gleiser, E.W.~Kolb and R.~Watkins,
Nucl. Phys. {\bf B 364},411 (1991); N.~Tetradis, Z. Phys. {\bf
C57}, 331 (1993).

\bibitem{2} M. Gleiser, Phys. Rev. Lett. {\bf 73}, 3495 (1994).

\bibitem{3} J. Borrill and M. Gleiser, Phys. Rev. D {\bf 51},
4111 (1995).

\bibitem{4} T. Shiromizu, M. Morikawa and J. Yokoyama,
Prog. Theor. Phys. {\bf 94}, 795 (1995).

\bibitem{5} M. Gleiser, A.F. Heckler and E.W. Kolb,
FERMILAB-Pub-95-374/A, DART-HEP-95/07, cond-mat/9512032.

\bibitem{6} M. Gleiser and A.F. Heckler, Phys. Rev. Lett {\bf 76}, 180
(1996). 

\bibitem{7} K. Enqvist, A. Riotto and I. Vilja, Phys. Rev. D {\bf
52},5556 (1995).

\bibitem{8} M. Dine, R.G. Leigh, P. Huet, A. Linde, D. Linde,
Phys. Rev. D {\bf 46}, 550 (1992).

\bibitem{9} P. Arnold and O. Espinosa, Phys. Rev. D {\bf 47}, 3546
(1993); W. Buchmuller, Z. Fodor, A. Hebecker, DESY-95-028,
hep-ph/9502321; K. Kajantie, M. Laine, K. Rummukainen,
M. Shaposhnikov, CERN-TH-95-263, hep-lat/9510020; P. Arnold, L. Yaffe,
Phys. Rev. D {\bf 49}, 3003 (1994); W. Buchmuller, O. Philipsen,
Nucl. Phys. {\bf B443}, 47 (1995); Z. Fodor and A. Hebecker,
Nucl. Phys. {\bf B432}, 127 (1994).

\bibitem{10} B.-H. Liu, L. McLerran, N. Turok, Phys. Rev. D {\bf 46},
2668 (1992).

\bibitem{10.5} A.F. Heckler, Phys. Rev. D {\bf 51}, 405 (1995).

\bibitem{11} P. Arnold, Phys. Rev. D {\bf 48}, 1539 (1993).

\bibitem{11.5} G.W. Anderson, L.J. Hall, Phys. Rev. D {\bf 45}, 2685
(1992). 

\bibitem{12} G.D. Moore, T. Prokopec, Phys. Rev. Lett {\bf 75}, 777
(1995); Phys. Rev. D {\bf 52}, 7182 (1993).

\bibitem{13} A.G. Cohen, D.B. Kaplan, A.E. Nelson,
Annu. Rev. Nucl. Part. Sci. {\bf 43}, 27 (1993).

\bibitem{14} A.G. Cohen, D.B. Kaplan, A.E. Nelson, Nucl. Phys. {\bf
B373}, 453 (1992); Phys. Rev. Lett {\bf 294}, 57 (1992).

\bibitem{15} V.A. Kuzmin, V.A. Rubakov, M.E. Shaposhnikov,
Phys. Lett. {\bf B155}, 36 (1985).

\bibitem{16} L. McLerran, et al., Phys. Lett. {\bf B256}, 451 (1991);
M. Dine, P. Huet, R. Singleton, Nucl. Phys. {\bf B375}, 625 (1992);
M. Dine et al., Phys. Lett. {\bf B257}, 351 (1991); A.G. Cohen,
A.E. Nelson, Phys. Lett. {\bf B297}, 111 (1992).

\bibitem{17} D. Comelli, M. Pietroni, A. Riotto, Phys. Lett. {\bf
B354}, 91 (1995).

\bibitem{18} A. Riotto, SISSA/AP/95/117, hep-ph/9510271.

\bibitem{19} A. Sopczak, Nucl. Phys. B (Proc. Suppl.) {\bf 37}, 168
(1995). 

\bibitem{19.5} G.F. Mazenko, O.T. Valls, and P. Ruggiero, Phys. Rev. B
{\bf 40}, 384(1989); M.C. Cross and P.C. Hohenberg,
Rev. Mod. Phys. {\bf 65}, 851 (1993); J. Armero, et al.,
Phys. Rev. Lett {\bf 76}, 3045 (1996) and references therein.

\bibitem{20} J. Langer, Ann. Phys. (N.Y.) {\bf 54}, 258 (1969).

\bibitem{21} A. Linde, Nucl. Phys. {\bf B216}, 421 (1983) [erratum:
{\bf B223}, 544 (1983)].


\bibitem{22} S. Coleman, Phys. Rev. D {\bf 15}, 2929 (1977);
C. Callan, S. Coleman, Phys. Rev. D {\bf 16}, 1762 (1977).

\bibitem{23} B. Link, Phys. Rev. Lett {\bf 68}, 2425 (1992).

\bibitem{24} M. Kamionkowski, K. Freese, Phys. Rev. Lett {\bf 69},
2743 (1992).

\bibitem{24.5} E.W. Kolb, M.S. Turner, {\it The Early Universe}
(Addison -- Wesley 1990), ch. 3.

\bibitem{25} P. Huet, K. Kajantie, R.G. Leigh, B.H. Liu, L. McLerran,
Phys. Rev. D {\bf 48}, 2477 (1993).

\bibitem{26} M. Abney, Phys. Rev. D {\bf 49}, 1777 (1994).

\bibitem{27} L. Rezzolla, SISSA-138/95/A (1995).

\bibitem{28} D.S. Lee, D. Boyanovsky, Nucl. Phys. {\bf B406}, 631
(1993); M. Morikawa, Phys. Rev. D {\bf 33}, 3607 (1986); B.L. Hu,
J. P. Paz, Y. Zhang in {\it The Origin of Structure in the Universe},
Ed. E. Gunzig, P. Nardone (Kluwer Acad. Publ. 1993).

\bibitem{29} M. Gleiser, R.O. Ramos, Phys. Rev. D {\bf 50}, 2441
(1994). 

\bibitem{30} M. Alford, M. Gleiser, Phys. Rev. D {\bf 48}, 2838
(1993). 

\bibitem{31} F. Alexander, S. Habib, Phys. Rev. Lett {\bf 71}, 955
(1993); M. Alford, H. Feldman, M. Gleiser, Phys. Rev. Lett {\bf 68},
1645 (1992).


\end{thebibliography}
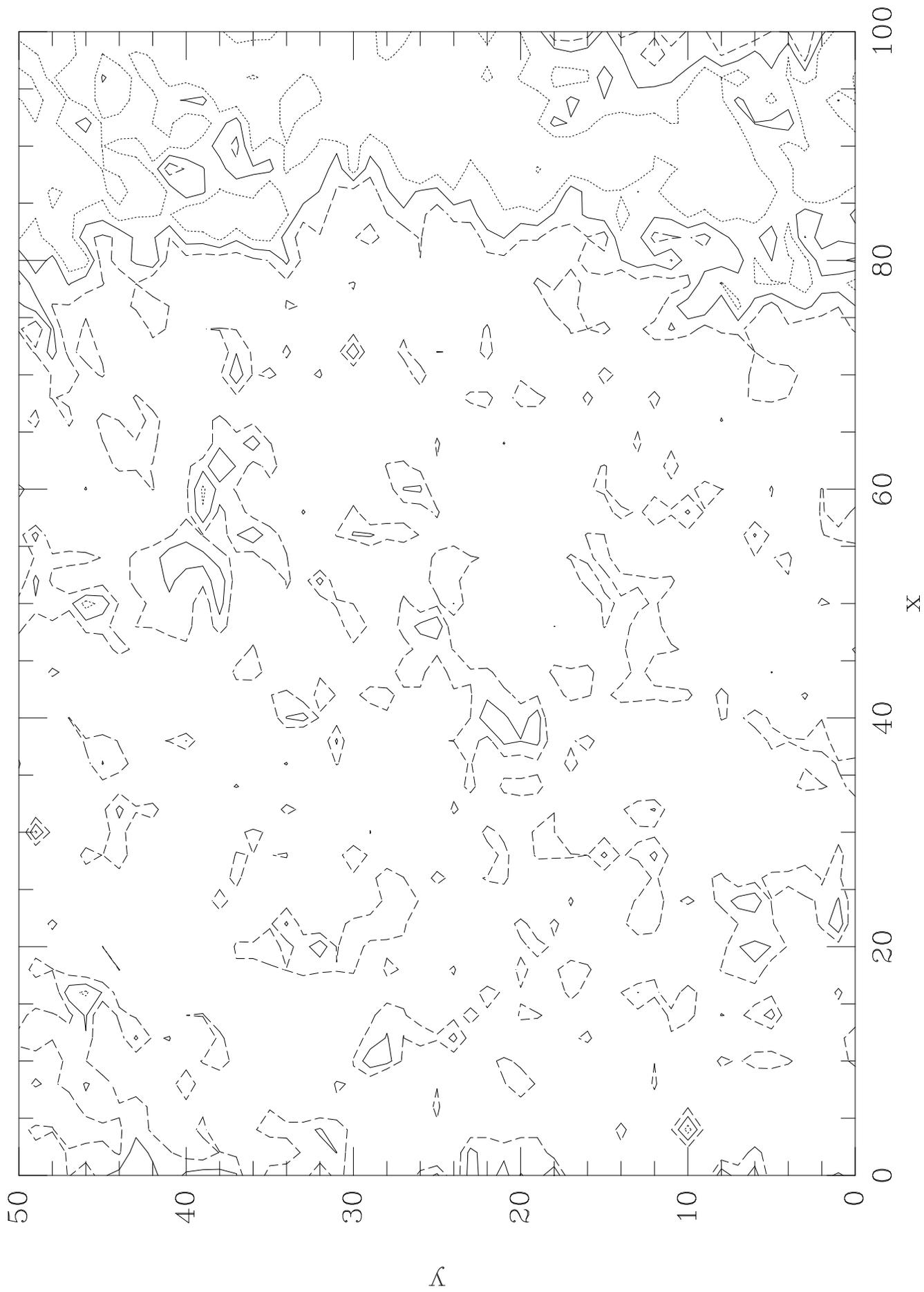
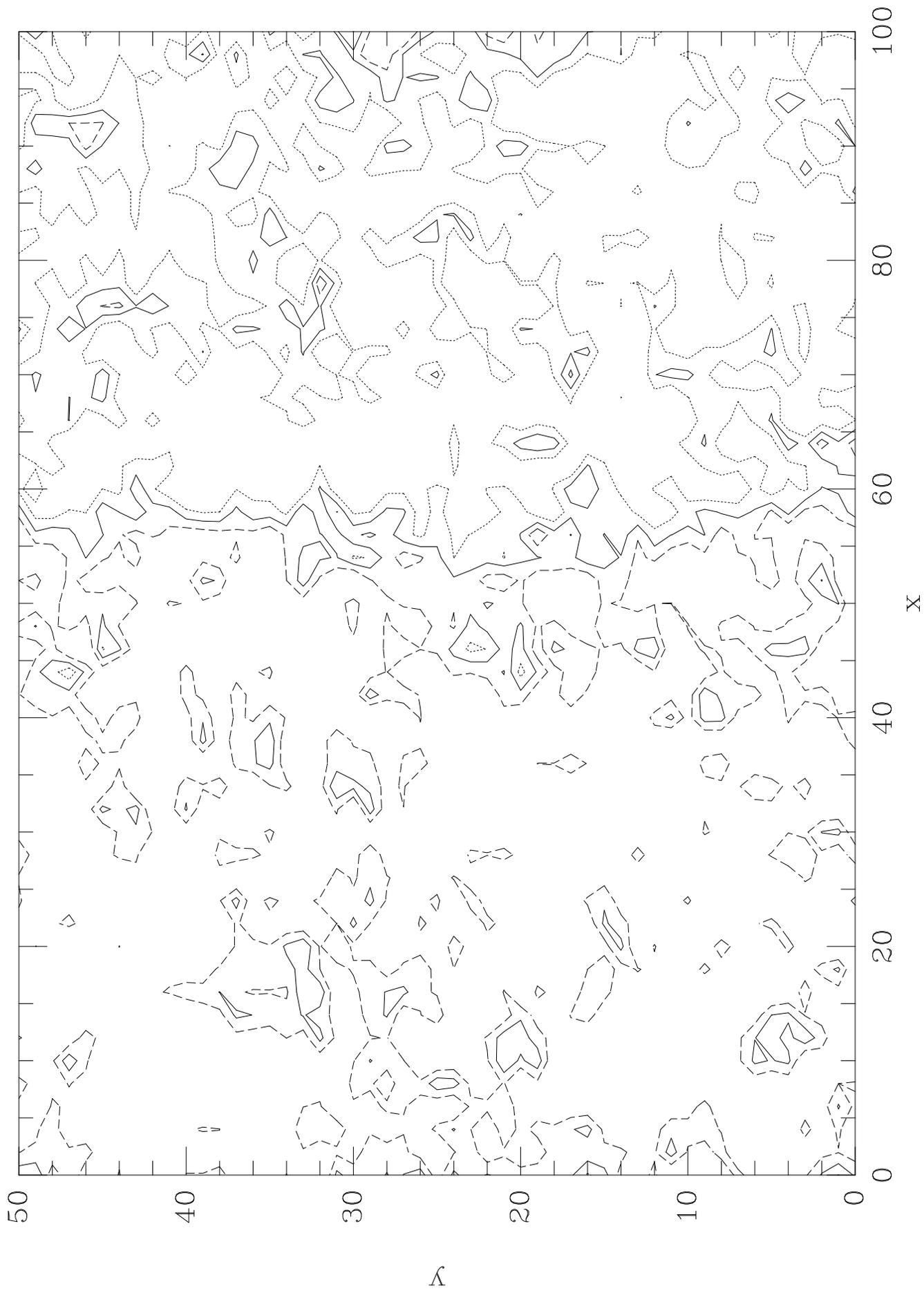
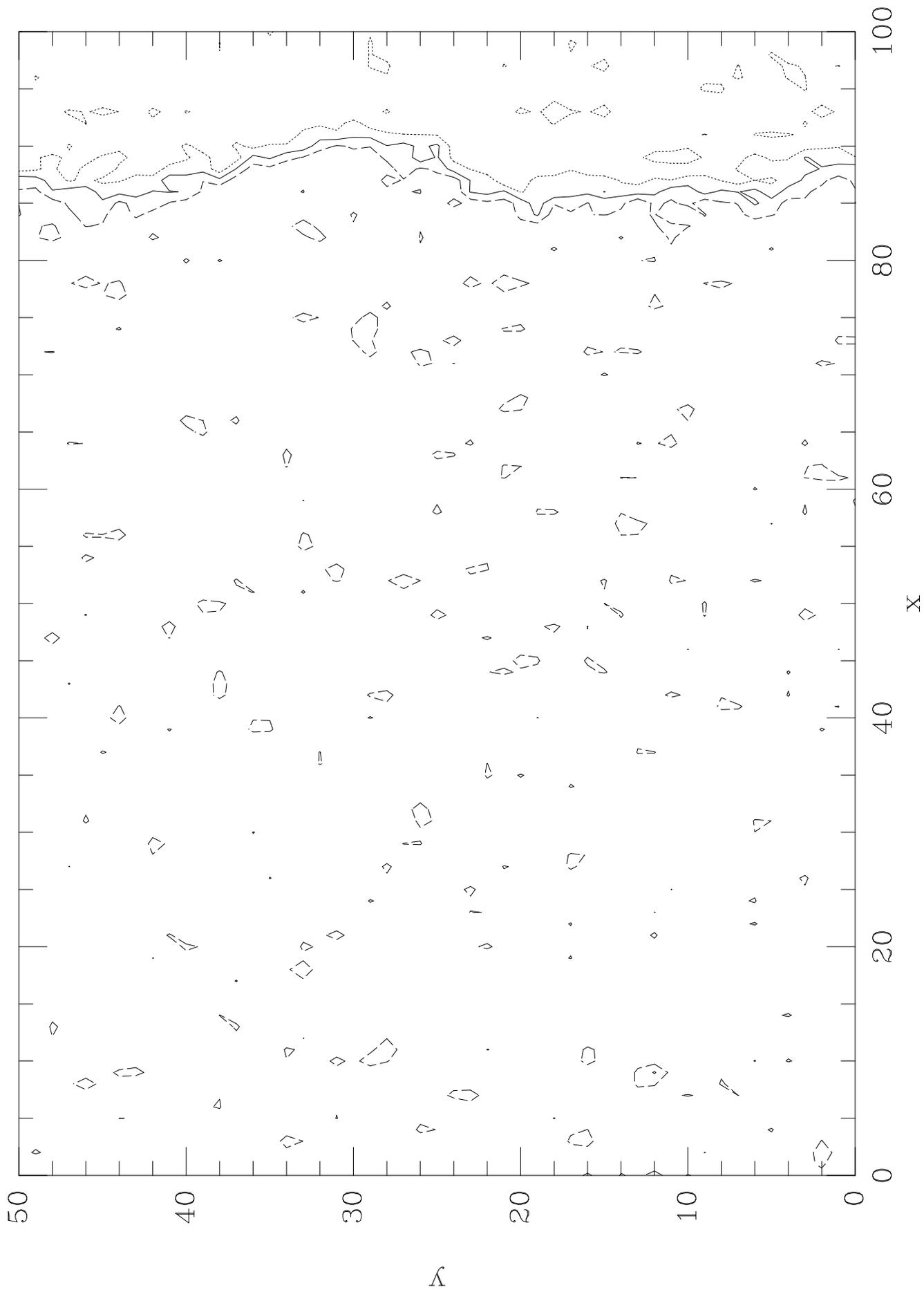
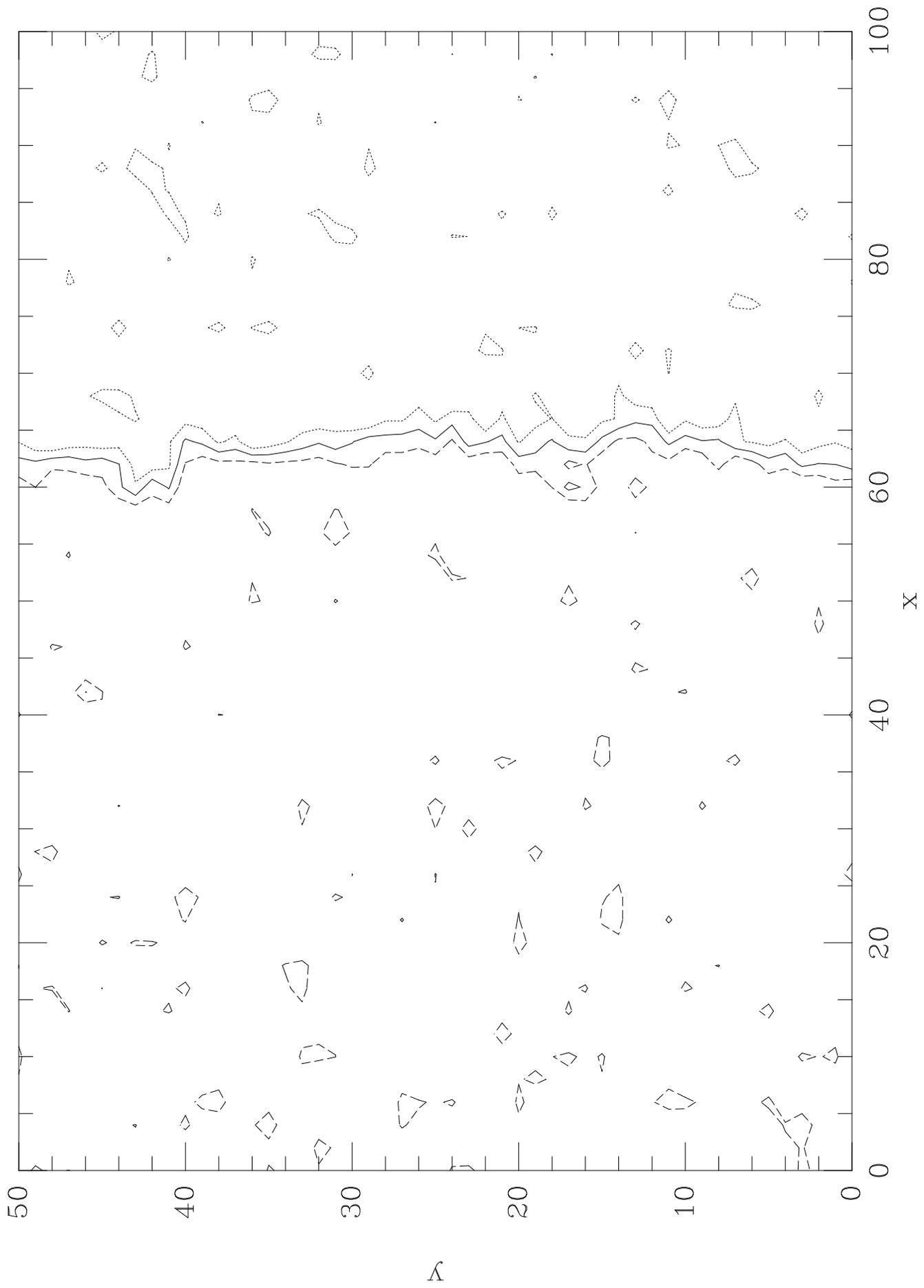
\end{document}